\documentclass[aps,prd,groupedaddress]{revtex4-1}

\usepackage{graphicx}
\usepackage{dcolumn}
\usepackage{bm}

\newcommand{\apjs}{\rm Astrophys.~J.~Supp.~}

\newcommand{\physrep}{\rm Phys.~Rep.~}

\begin{document}


\title{Constraints on the multi-lognormal magnetic fields from the observations of the cosmic microwave background and the matter power spectrum}

\author{Dai G. Yamazaki$^{1}$}
\author{Kiyotomo Ichiki$^{2}$}
\author{Keitaro Takahashi$^{3}$}
 \email{yamazaki.dai@nao.ac.jp}
\affiliation{%
$^{1}$National Astronomical Observatory of Japan, Mitaka, Tokyo 181-8588, Japan}%
\affiliation{%
$^{2}$
Kobayashi-Maskawa Institute for the Origin of Particles and the Universe, Nagoya University,
Furo-cho, Chikusa-ku, Nagoya, Aichi, 464-8602, Japan
}%
\affiliation{%
$^{3}$
Faculty of Science, Kumamoto University, 2-39-1,
Kurokami, Kumamoto, Kumamoto, 860-8555, Japan
}%
\date{\today}

\begin{abstract}
Primordial magnetic fields (PMFs), which were generated in the early Universe before recombination, affect the motion of plasma and then the cosmic microwave background and the matter power spectrum. We consider constraints on PMFs with a characteristic correlation length from the observations of the anisotropies of the cosmic microwave background and the matter power spectrum. The spectrum of PMFs is modeled with multi-lognormal distributions, rather than power-law distribution, and we derive constraints on the strength $|\mathbf{B}_k|$ at each wave number $k$ along with the standard cosmological parameters in the flat Universe and the foreground sources. We obtain upper bounds on the field strengths at $k=10^{-1}, 10^{-2}, 10^{-4}$,  and $10^{-5}$ Mpc$^{-1}$ as 4.7 nG, 2.1 nG, 5.3 nG,  and 10.9 nG ($2\sigma$ C.L.) respectively, while the field strength at $k=10^{-3} $Mpc$^{-1}$  turns out to have a finite value as $|\mathbf{B}_{k = 10^{-3}}| =  6.2 \pm 1.3 $ nG ($1\sigma$ C.L.). This finite value is attributed to the finite values of BB mode data at $\ell > 300$ obtained from the QUAD experiment. If we do not include the BB mode data, we obtain only the upper bound on $B_{k=10^{-3}}$.
\end{abstract}
\maketitle 
\section{\label{sec:introduction}Introduction}
From discoveries of magnetic fields in clusters of galaxies \cite{2004IJMPD..13.1549G,Wolfe:1992ab,Clarke:2000bz,Xu:2005rb}, many authors have studied cosmological magnetic fields.
The origins of primordial magnetic fields (PMFs) have been studied by many authors \cite{Turner:1987bw,Ratra:1991bn,Bamba:2004cu, Vachaspati:1991nm,Kibble:1995aa,Ahonen:1997wh,Joyce:1997uy, Takahashi:2005nd,Hanayama:2005hd,ichiki:2006sc,2011PhR...505....1K}.
PMFs may manifest themselves in the temperature and polarization anisotropies of the cosmic microwave background (CMB) \cite{Subramanian:1998fn,Jedamzik:1999bm,Mack:2001gc,Subramanian:2002nh,Lewis:2004ef,Yamazaki:2004vq,Kahniashvili:2005xe,Challinor:2005ye,Dolgov:2005ti,Gopal:2005sg,Yamazaki:2005yd,Kahniashvili:2006hy,Yamazaki:2006bq,Yamazaki:2006ah,Giovannini:2006kc,Yamazaki:2007oc,Paoletti:2008ck,Yamazaki:2008bb,2008nuco.confE.239Y,Sethi:2008eq,Kojima:2008rf,2008PhRvD..78f3012K,Giovannini:2008aa,2010PhRvD..81b3008Y,2010PhRvD..81j3519Y,2010AdAst2010E..80Y,2010PhRvD..82h3005K,2012PhR...517..141Y,2012PhRvD..86d3510S}, the large-scale structure \cite{Sethi:2003vp,Sethi:2004pe,Yamazaki:2006mi,2013ApJ...770...47K}, and other physical phenomena \cite{1995APh.....3...95G,1996PhLB..379...73G,1996PhRvD..54.7207K,Jedamzik:1996wp,Grasso:2000wj,Banerjee:2004df,Shaw:2009nf,2012arXiv1204.6164K,2012PhRvD..86l3006Y}.
The effects of PMFs on physics in the early Universe and constraints on the PMFs from the cosmological observations are well-investigated topics.

A power-law spectrum has often been considered for the study of PMFs in the literature (see Refs. \cite{Grasso:2000wj,2011PhR...505....1K,2012PhR...517..141Y,2012PhRvD..86l3006Y} and references therein) because it is expected for PMFs generated by inflation. On the other hand, causal processes such as bubble collisions during phase transition would generate PMFs with a characteristic scale. To study such PMFs, in our previous work \cite{2011PhRvD..84l3006Y}, we considered PMFs with a lognormal distribution (LND), which is parametrized by the characteristic wave number $k$, the field strength at the wave number $|\mathbf{B}_{k}|$ and the width $\sigma_\mathrm{M}$ of the spectrum. We derived constraints on the parameters from the observations of anisotropies of the CMB fixing cosmological parameters to the WMAP best-fit values. This type of PMF spectrum is also useful to study which scale of PMFs mainly affects the CMB.

Recently, ACT \cite{2011ApJ...739...52D} and SPT \cite{2011ApJ...743...28K} projects published more precise results of the CMB observations on small scales with $\ell > 1000$. Because the effect of PMFs is relatively strong at smaller scales (see Refs. \cite{Grasso:2000wj,2011PhR...505....1K,2012PhR...517..141Y,2012PhRvD..86l3006Y} and references therein), constraints on PMFs are expected to become stronger with these data. However, since the Sunyaev-Zel'dovich (SZ) effect by galaxy clusters and emissions from radio galaxies also contributes to the anisotropies of the CMB on these scales, the PMF parameters would have degeneracies with the amplitudes of these foreground effects. Therefore, in this paper, extending our previous work, we derive constraints on the PMF parameters varying cosmological and foreground parameters as well, to obtain more reliable constraints from the CMB and the matter power spectrum (MPS) observations. We also consider, as the spectrum of the PMFs, a multi-lognormal distribution with five characteristic wave numbers.

We introduce the effects of the PMFs on the CMB and the MPS in Sec. 2. In Sec. 3 we illustrate the constraint on $|\mathbf{B}_{k = 10^{n}}|$ at each $k_\mathrm{M}$ from the CMB and the MPS and discuss the degeneracies between the constrained $|\mathbf{B}_{k = 10^{n}}|$ and cosmological parameters. We summarize our results and describe future plans in Sec. 4.
\section{\label{sec:models}Model and Method}
In this section, we show how to consider the effects of the PMF on the CMB and MPS. 
Before the recombination of protons and electrons, ionized baryons are directly affected by the PMF through the Lorentz force. 
Since the photons and the baryons are tightly coupled before the last scattering of photons, the PMF indirectly affects the photons. 
Also, the PMFs indirectly affect the cold dark matter (CDM) through gravitational interaction.
We have an assumption that the PMF is produced at some time during the radiation-dominated era. 
Since the cosmological magnetic fields are observed with amplitudes of the orders of 1-10 nG, we also assume that the field strengths of the PMFs, $|\mathbf{B}_{k = 10^{n}}|$s, are less than 10 nG.
In this case, since $|\mathbf{B}_{k = 10^{n}}|$s are the average strength of the PMFs and the energy density of the PMFs $(|\mathbf{B}_{k = 10^{n}}|=10\mathrm{nG})^2/(8\pi) \sim 10^{-5}\times \rho_\gamma)$ is much smaller than that of the background photons, 
we can treat the PMF energy density as a first-order perturbation with the flat Friedmann-Robertson-Walker as the background spacetime, and all of the back reactions from the fluid to the PMF can be neglected (see Ref. \cite{Mack:2001gc} for details).

We modify the CAMB code \cite{camb}, taking into consideration the PMF effects. 
We use the adiabatic initial condition for the time evolution of the CMB and the MPS with the PMFs. Details of these are summarized in Ref.~\cite{2011PhRvD..84l3006Y}. 
We use the spectrum of the PMFs as
\begin{eqnarray}
f_\mathrm{LND}(k;k_\mathrm{M}, \sigma_\mathrm{M})
=
\frac{1}{k \sigma_\mathrm{M} \sqrt{2\pi}}
	\exp \left\{
			-\frac{
				\left[
					\ln{(k)}-\ln{(k_\mathrm{M})}
				\right]^2
				}
				{2\sigma_\mathrm{M}^2}
		 \right\}
\label{eq:power_lg},
\end{eqnarray}
where $k_\mathrm{M}$ is the characteristic scale depending on the PMF
generation model and $\sigma_\mathrm{M}$ is the scale parameter.
The detailed mathematical description of this spectrum is also defined in Ref.~\cite{2011PhRvD..84l3006Y}.
Since our goal is qualitatively understanding how the PMFs are limited scale by scale, for simplicity, we fix the scale parameter at $\sigma_\mathrm{M} = 1$.

In this paper, we limit the PMF strengths for each characteristic scale together with the other cosmological parameters from the CMB and the MPS using a Markov chain Monte Carlo (MCMC) method \cite{cosmomc}.
We constrain five strengths of the PMF at $\log_{10}(k_\mathrm{M}\mathrm{[Mpc^{-1}]}) =$ -1, -2, -3, -4, and -5, which we denote as $|\mathbf{B}_{k = 10^{-1}}|$, $|\mathbf{B}_{k = 10^{-2}}|$, $|\mathbf{B}_{k = 10^{-3}}|$, $|\mathbf{B}_{k = 10^{-4}}|$, and $|\mathbf{B}_{k = 10^{-5}}|$, respectively, and 15 cosmological parameters, i.e.,~
$
[
\Omega_b h^2,
$
$
	\Omega_c h^2,
$
$
	\theta,
$
$
	\tau_C,
$
$
	n_s,
$
$
	\log(10^{10}A_s),
$
$
	A_t/A_s, 
$
$
	A^\mathrm{LEN},
$
$
	A^\mathrm{CL}, 
$
$
	A^\mathrm{PS},
$
$
	A^\mathrm{SZ}_\mathrm{MAP},
$
$
	A^\mathrm{SZ}_\mathrm{ACB},
$
$
	A^\mathrm{SZ}_\mathrm{QUD},
$
$
	A^\mathrm{SZ}_\mathrm{ACT},
$
$
	A^\mathrm{SZ}_\mathrm{SPT}
],
$
where
$\Omega_c h^2$
and
$\Omega_b h^2$
are the CDM and baryon densities in units of the critical density,
$h$ denotes the Hubble parameter in units of 100 km s$^{-1}$Mpc$^{-1}$,
$\theta$ is the ratio of the sound horizon to the angular diameter distance,
$\tau_C$
is the optical depth for Compton scattering,
$n_s$
is the spectral index of the primordial scalar fluctuations,
$A_s$
is the amplitude of primordial scalar fluctuations,
$A_t$
is the amplitude of the primordial tensor fluctuations,
$A^\mathrm{LEN}$ is the amplitude of the weak lensing,
$A^\mathrm{CL}$ and $A^\mathrm{PS}$ are the amplitudes of cluster point sources and Poisson point sources at $\ell = 3000$,
 and 
$A^\mathrm{SZ}_\mathrm{X}$ is the SZ effect amplitudes in observations (X), which are denoted with subscripts WMAP (MAP) \cite{WMAP_9yr_Arxiv}, ACBAR (ACB) \cite{Kuo:2006ya}, QUAD (QUD) \cite{2009ApJ...705..978B}, ACT \cite{2011ApJ...739...52D}, and SPT \cite{2011ApJ...743...28K}.
Note that each SZ spectrum model is published in LAMBDA \cite{Note1}, which is the spectrum based on  Ref.~\cite{Komatsu:2002wc}.
We fix the spectral index of the primordial tensor fluctuations as $n_t =-(A_s/A_t)/8 $.
For all the cosmological parameters, we use the same priors as those adopted in the WMAP analysis \cite{WMAP_9yr_Arxiv}.
We use the CMB and the MPS observation data
 sets as follows: 
WMAP\cite{WMAP_9yr_Arxiv}, ACBAR\cite{Kuo:2006ya}, QUAD\cite{2009ApJ...705..978B}, ACT\cite{2011ApJ...739...52D}, SPT\cite{2011ApJ...743...28K}, SDSS\cite{Tegmark:2006az}.
\section{\label{sec:results}Results and Discussions}
Our MCMC algorithm is performed until all of the parameters including the PMF strengths are well converged to the values listed in Table 1. We find that the minimum total $\chi^2$ improves from 12563.2 to 12554.9 by introducing the PMFs. As we see below, this improvement mainly comes from the finite B mode spectrum at $k = 10^{-3} \mathrm{Mpc^{-1}}$ in the QUAD experiment.

\begin{figure}
\includegraphics[width=0.8\textwidth]{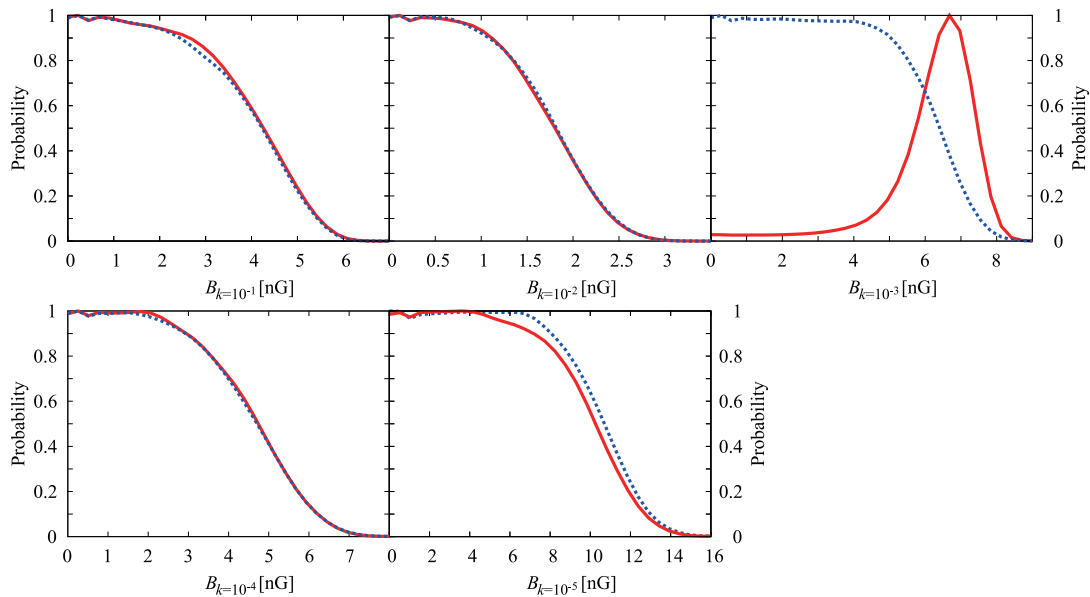}
\caption{\label{fig1}
Probability distributions of the PMF strengths at $k_\mathrm{M}~=~10^{-5},~10^{-4},~10^{-3},~10^{-2},$ and $10^{-1}$ Mpc$^{-1}$ from the CMB and MPS data with the BB mode (solid) and without (dotted).
} 
\end{figure}

Figure \ref{fig1} shows the probability distribution functions of PMF strengths obtained by our MCMC analysis. For the strengths except the one at $k = 10^{-3} \mathrm{Mpc^{-1}}$, upper bounds are obtained at $2\sigma$ as follows:
\begin{eqnarray}
|\mathbf{B}_{k = 10^{-1}}| &<& 4.685~\mathrm{nG}~\mathrm{at}~k = 10^{-1} \mathrm{Mpc^{-1}}, \\
|\mathbf{B}_{k = 10^{-2}}| &<& 2.090~\mathrm{nG}~\mathrm{at}~k = 10^{-2} \mathrm{Mpc^{-1}}, \\
|\mathbf{B}_{k = 10^{-4}}| &<& 5.310~\mathrm{nG}~\mathrm{at}~k = 10^{-4} \mathrm{Mpc^{-1}} ~\mathrm{and} \\
|\mathbf{B}_{k = 10^{-5}}| &<& 10.91~\mathrm{nG}~\mathrm{at}~k = 10^{-5} \mathrm{Mpc^{-1}}.
\end{eqnarray}
On the other hand, for the strength at $k = 10^{-3} \mathrm{Mpc^{-1}}$, a nonzero value is favored at more than $2.5\sigma$:
\begin{eqnarray}
4.867~\mathrm{nG} < |\mathbf{B}_{k = 10^{-3}}| < 7.491~\mathrm{nG}~ (1\sigma).
\end{eqnarray}
In Fig. \ref{fig1}, the probability distribution functions obtained without CMB BB modes (curl-type polarization fluctuations) are also shown. We can see, in this case, that the lower bound of the PMF strength at $k = 10^{-3} \mathrm{Mpc^{-1}}$ disappears. Thus, the nonzero value is favored by the BB mode data. We shall explain and discuss these results below.

\begin{table}
\begin{tabular}{ccc} 
\multicolumn{3}{c}{\it Cosmological Parameters}\\
\hline
\multicolumn{1}{c}{Parameter} &
\multicolumn{1}{c}{mean}&
\multicolumn{1}{c}{best fit}\\
\hline
$\Omega_b h^2$ &
$0.02279 \pm 0.00052$ &
$0.02258$ \\
$\Omega_c h^2$ &
$0.1104\pm 0.0046$ &
$0.1118$ \\
$\theta$ &
$1.042\pm 0.0015$ &
$1.043$ \\
$\tau_C$ &
$0.08687\pm 0.0132$ &
$0.08126$ \\
$n_s$ &
$0.9597\pm 0.0120$ &
$0.9592$ \\
$\ln(10^{10}A_s)$ &
$3.172\pm 0.047$ &
$3.174$\\
$A_t/A_s$ &
$< 0.06047 ( 68\% \mathrm{C.L.}), < 0.1528 ( 95\% \mathrm{C.L.})$ &
$0.01106$\\
$A^\mathrm{LEN}$ &
$0.9141\pm 0.228$ &
$ 0.874$\\
$A^\mathrm{PS}$ &
$15.24\pm 2.01$ &
$16.37$\\
$A^\mathrm{CL}$ &
$< 7.254( 68\% \mathrm{C.L.}), < 11.94( 95\% \mathrm{C.L.})$ &
$7.535$ \\
$A^\mathrm{SZ}_\mathrm{MAP}$ &
$< 2.762( 68\% \mathrm{C.L.}), < 5.114( 95\% \mathrm{C.L.})$ &
$0.3509$ \\
$A^\mathrm{SZ}_\mathrm{ACB}$ &
$< 3.468 ( 68\% \mathrm{C.L.}), < 6.644( 95\% \mathrm{C.L.})$ &
$1.447$ \\
$A^\mathrm{SZ}_\mathrm{QUD}$ &
$< 2.550( 68\% \mathrm{C.L.}), < 5.448( 95\% \mathrm{C.L.})$ &
$0.7729$ \\
$A^\mathrm{SZ}_\mathrm{ACT}$ &
$< 0.2943( 68\% \mathrm{C.L.}), < 0.7207( 95\% \mathrm{C.L.})$ &
$0.09135$ \\
$A^\mathrm{SZ}_\mathrm{SPT}$ &
$< 8.405 ( 68\% \mathrm{C.L.}), < 12.78( 95\% \mathrm{C.L.})$ &
$6.102$ \\
$|\mathbf{B}_{k = 10^{-1}}|\mathrm{(nG)}$ &
$< 2.921 ( 68\% \mathrm{C.L.}), < 4.685 ( 95\% \mathrm{C.L.})$ &
$1.300$\\
$|\mathbf{B}_{k = 10^{-2}}|\mathrm{(nG)}$ &
$< 1.257 ( 68\% \mathrm{C.L.}), < 2.090 ( 95\% \mathrm{C.L.})$ &
$0.410$\\
$|\mathbf{B}_{k = 10^{-3}}|\mathrm{(nG)}$ &
$\mathbf{6.179} \pm \mathbf{1.312}$ &
$6.728$\\
$|\mathbf{B}_{k = 10^{-4}}|\mathrm{(nG)}$ &
$< 3.253 ( 68\% \mathrm{C.L.}), < 5.310 ( 95\% \mathrm{C.L.})$ &
$0.465$\\
$|\mathbf{B}_{k = 10^{-5}}|\mathrm{(nG)}$ &
$< 6.992 ( 68\% \mathrm{C.L.}), < 10.91 ( 95\% \mathrm{C.L.})$ &
$1.766$\\
\hline
\hline
\end{tabular}
\caption{Confidence intervals ($1\sigma$) and upper bounds ($2\sigma$) on strengths of PMFs, $\Lambda$CDM model parameters, and foreground parameters from a fit to WMAP\cite{WMAP_9yr_Arxiv}, ACBAR\cite{Kuo:2006ya}, QUaD\cite{2009ApJ...705..978B}, ACT\cite{2011ApJ...739...52D}, SPT\cite{2011ApJ...743...28K}, SDSS\cite{Tegmark:2006az} data.}
\label{CMBonlyTable}
\end{table}

\begin{figure}
\includegraphics[width=1.0\textwidth]{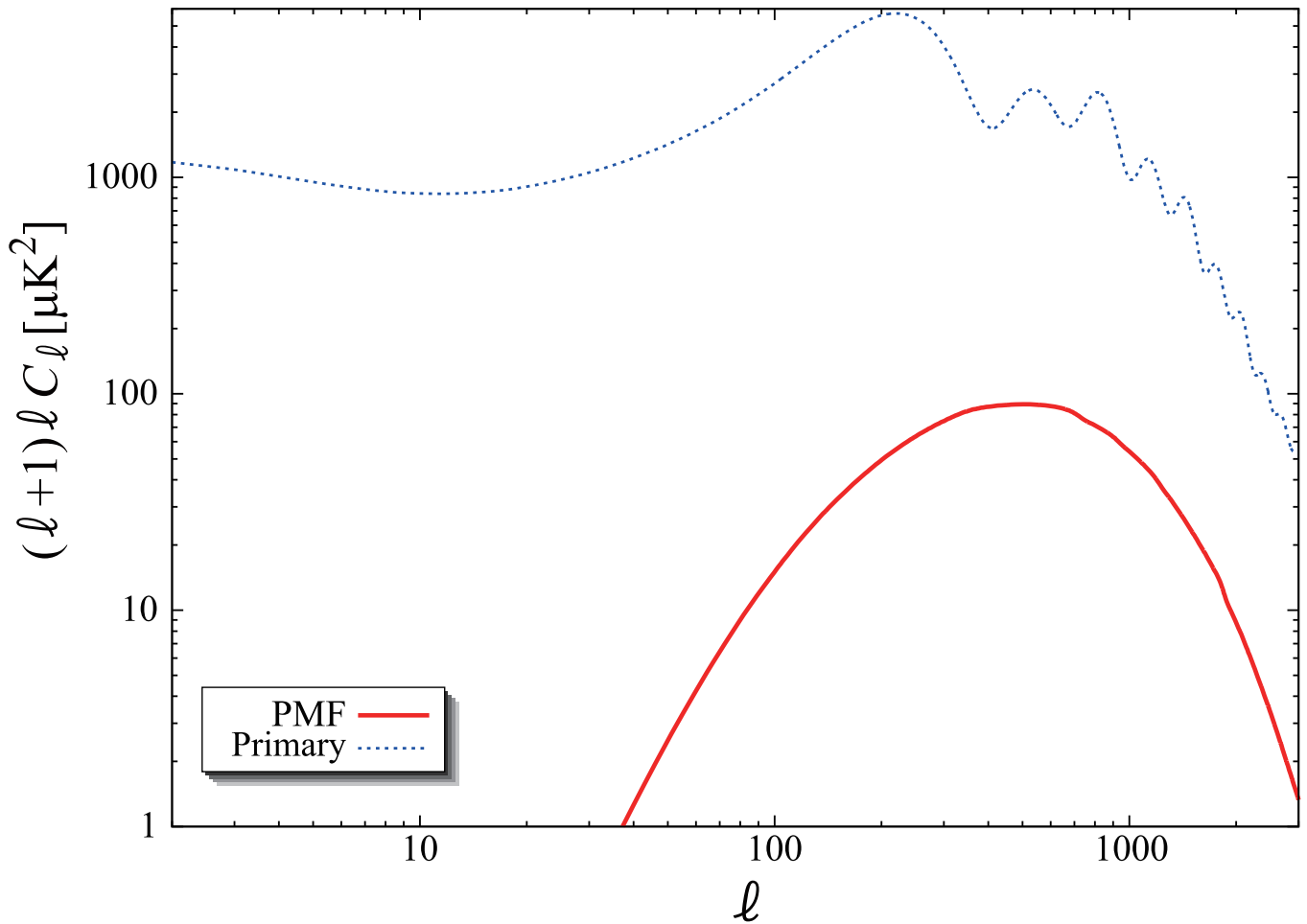}
\caption{\label{fig2}
CMB temperature spectra from the primary fluctuations (with foreground sources) and PMF with the best-fit parameters and $\sigma_\mathrm{M}=1.0$.
}
\end{figure}

\begin{figure}
\includegraphics[width=1.0\textwidth]{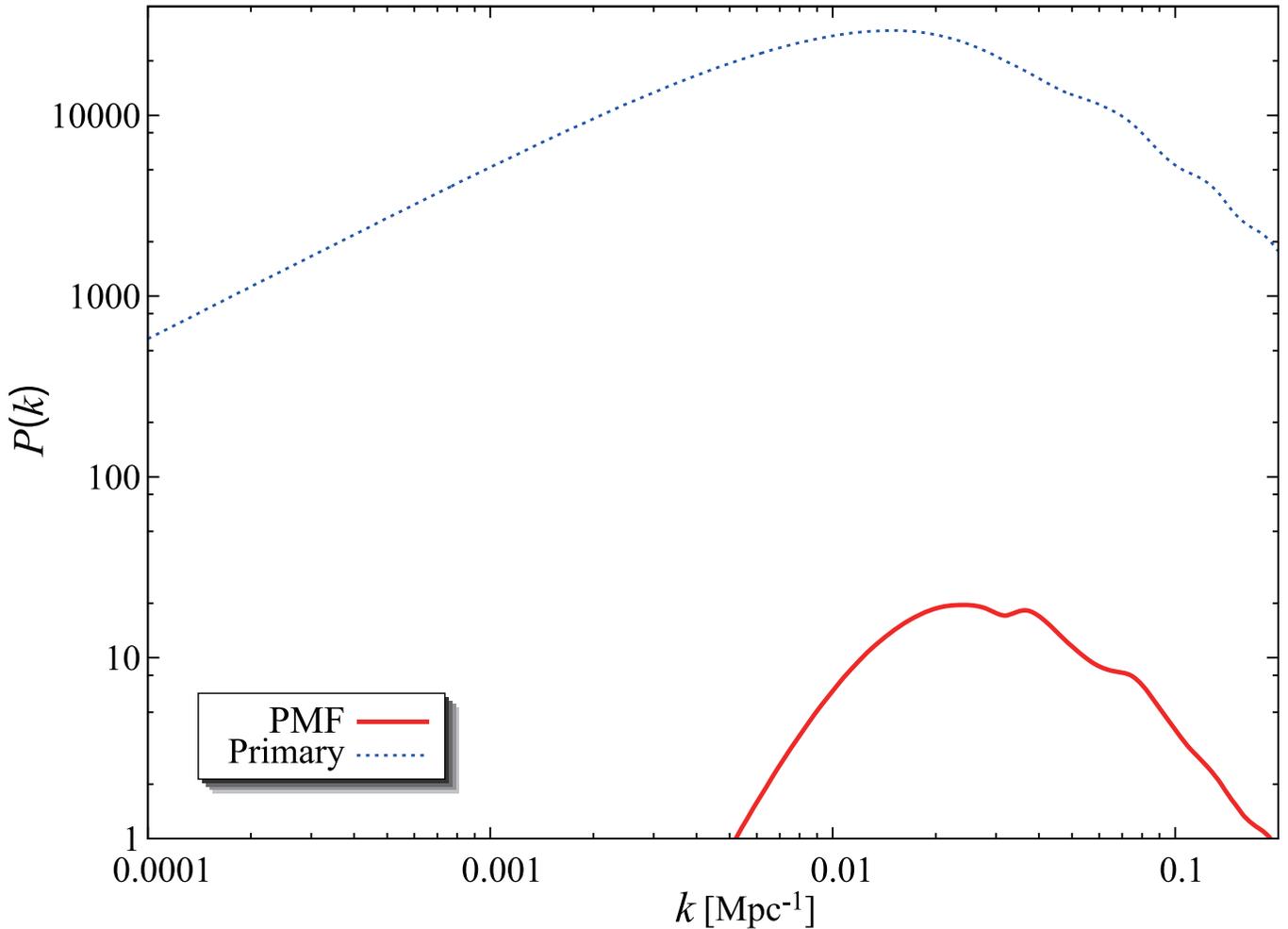}
\caption{\label{fig3}
Matter power spectra from the primary fluctuations and PMF with the best-fit parameters and $\sigma_\mathrm{M}=1.0$.
} 
\end{figure}

Figures \ref{fig2} and \ref{fig3} show the CMB temperature spectra and the MPS calculated with the best-fit parameters obtained above. We find the CMB temperature anisotropies due to the PMFs is a few percent of the primary fluctuations with foreground sources (lensing, CL, PS, and SZ effects) around the peaks and are comparable with the primary and foreground spectrum on $\ell <$ 1700. On the other hand, the contribution of PMFs to MPS is at most $0.05\%$ of the primary fluctuations. Actually, constraints on the PMF strengths mainly come from the CMB data.

If the amplitudes of CMB spectra from the PMFs except $k=10^{-3}$Mpc$^{-1}$ are comparable to the amplitude of the primary spectrum, the shape of the total spectrum of CMB (LND + primary) differs substantially from the observation results\cite{2011PhRvD..84l3006Y}. On the other hand, from Fig. \ref{fig2} and Ref. \cite{2011PhRvD..84l3006Y}, the peak of the CMB spectrum from the PMFs at $k=10^{-3}$Mpc$^{-1}$ is located on 200 $< \ell <$ 300, and this spectrum around the peak has a similar slope to the observed one. Therefore, only the PMFs at $k=10^{-3}$Mpc$^{-1}$ can refine the TT CMB spectrum.

The margins of errors of the observation on 300 $< \ell <$ 1700 are comparable in the differences between the theoretical CMB with the PMFs and the primary ones (without the PMFs). Therefore, the amplitude of the PMF is mainly constrained by the observational data on 300 $< \ell <$ 1700.

\begin{figure}
\includegraphics[width=1.0\textwidth]{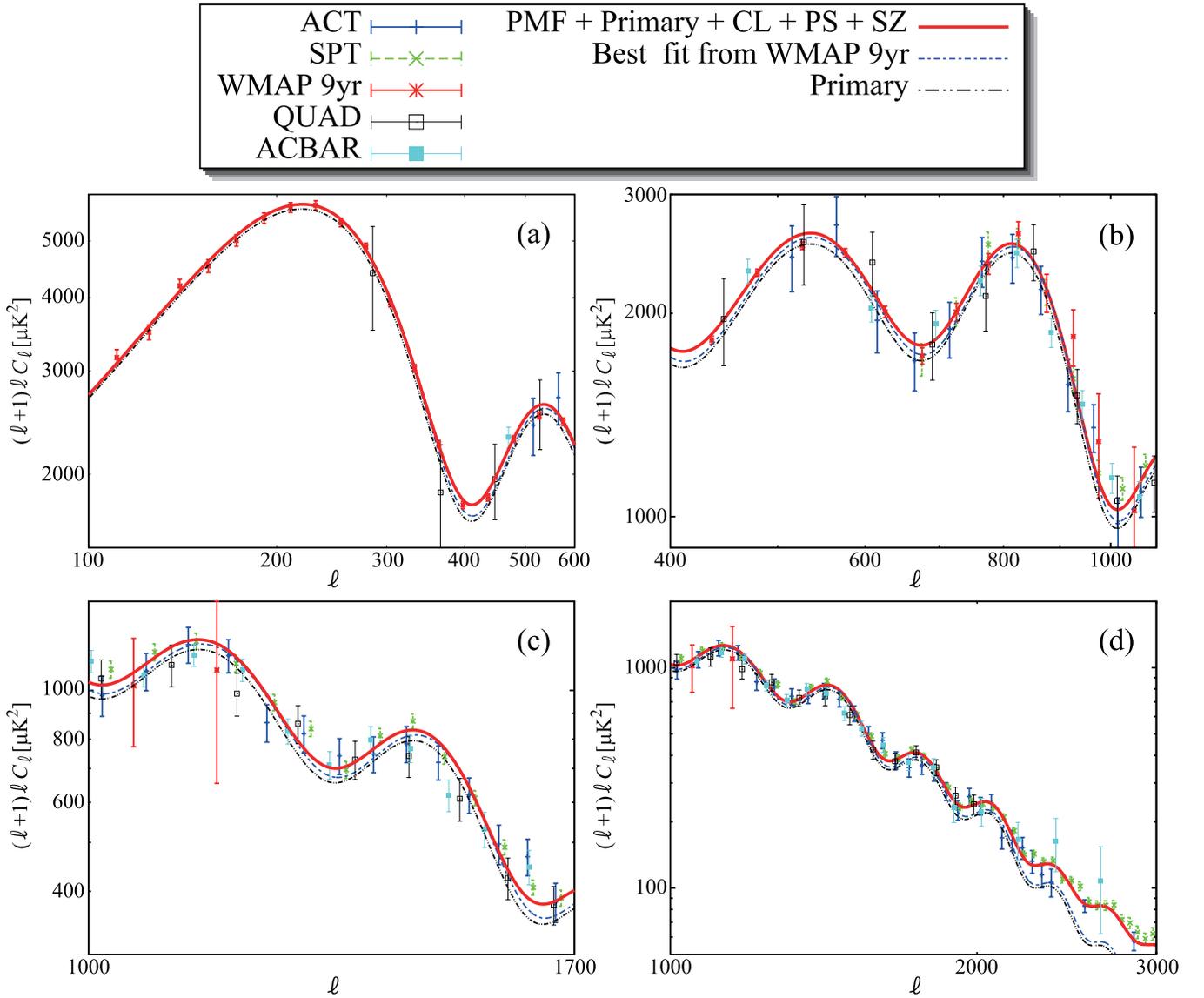}
\caption{\label{fig4}
Comparison of the best-fit total CMB power spectrum of the TT mode including the PMF with the primary one.  
Plots show various  multipole ranges for (a)  $100 < \ell < 600$ , (b)  $400 < \ell < 1100$, (c) $1000 < \ell < 1700$, and (d) $1000 < \ell < 3000$.
Curves and dots with error bars in all panels are the theoretical lines and observation results as indicated in the legend box on the figure.
The best fit from WMAP 9 year data includes neither the LND nor foreground contributions.} 
\end{figure}

From Fig.~\ref{fig4}, the effect of the PMFs, which is dominated by the LND-PMF spectrum at $k=10^{-3}$ Mpc$^{-1}$, is not dominant on $\ell > $ 2000, while the SZ effects and the point source contribution from clusters and radio galaxies dominant on these scale. Therefore, the PMFs do not have degeneracies with foreground components\cite{Note2}.

\begin{figure}
\includegraphics[width=1.0\textwidth]{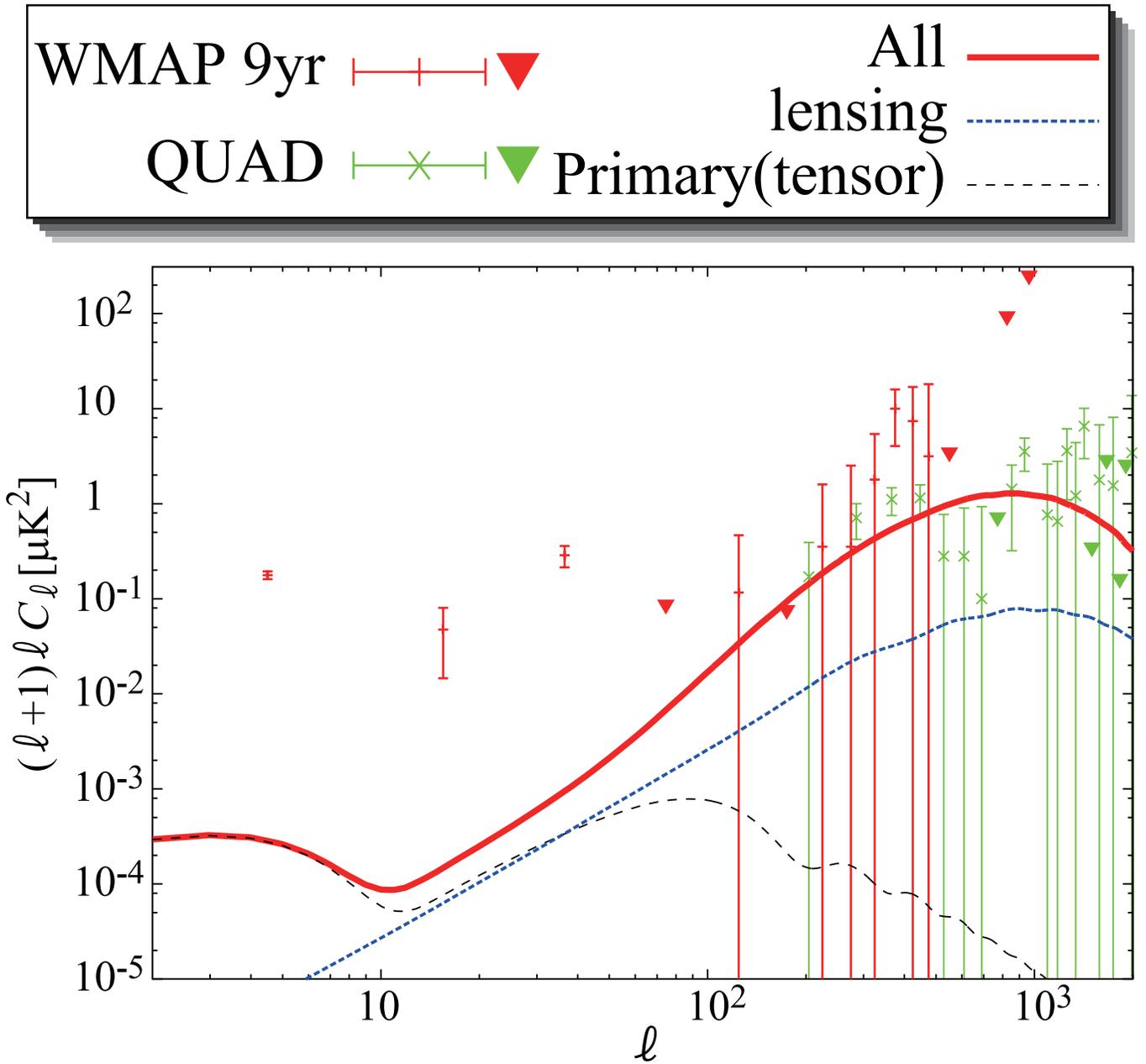}
\caption{\label{fig5}
Comparison of the best-fit total CMB power spectrum of the BB mode including the PMF with the observed one. Curves and dots with error bars are the theoretical lines and observation results as indicated in the legend box on the figure. Downward arrows for the error bars in this figure indicate that the data points are upper limits. The tensor-to-scalar ratio ($A_t/A_s$) of the primary BB mode is 0.01106.
} 
\end{figure}

The most striking effect of PMFs can be seen in the BB mode spectrum. Figure \ref{fig5} illustrates a comparison between the BB mode spectra with and without the best-fit LND-PMF spectrum. From this figure, we find that the BB mode spectrum with best-fit parameters is dominated by the PMF at small scales, and the BB mode data obtaind by the QUAD experiment is fitted better with the PMF than without it. This is why the nonzero PMF strength is favored when adding BB mode data as seen in Fig. \ref{fig1}. However, it should be noted that the current BB mode observations would have relatively large systematic and observational errors. Thus, although we need more precise observations of BB modes by future projects such as Planck and POLARBEAR to obtain a more solid conclusion, the current data imply the excess amplitude in BB mode, which can be well explained by the existence of PMF with a characteristic wave number of $k=10^{-3}$Mpc$^{-1}$.

\begin{figure}
\includegraphics[width=1.0\textwidth]{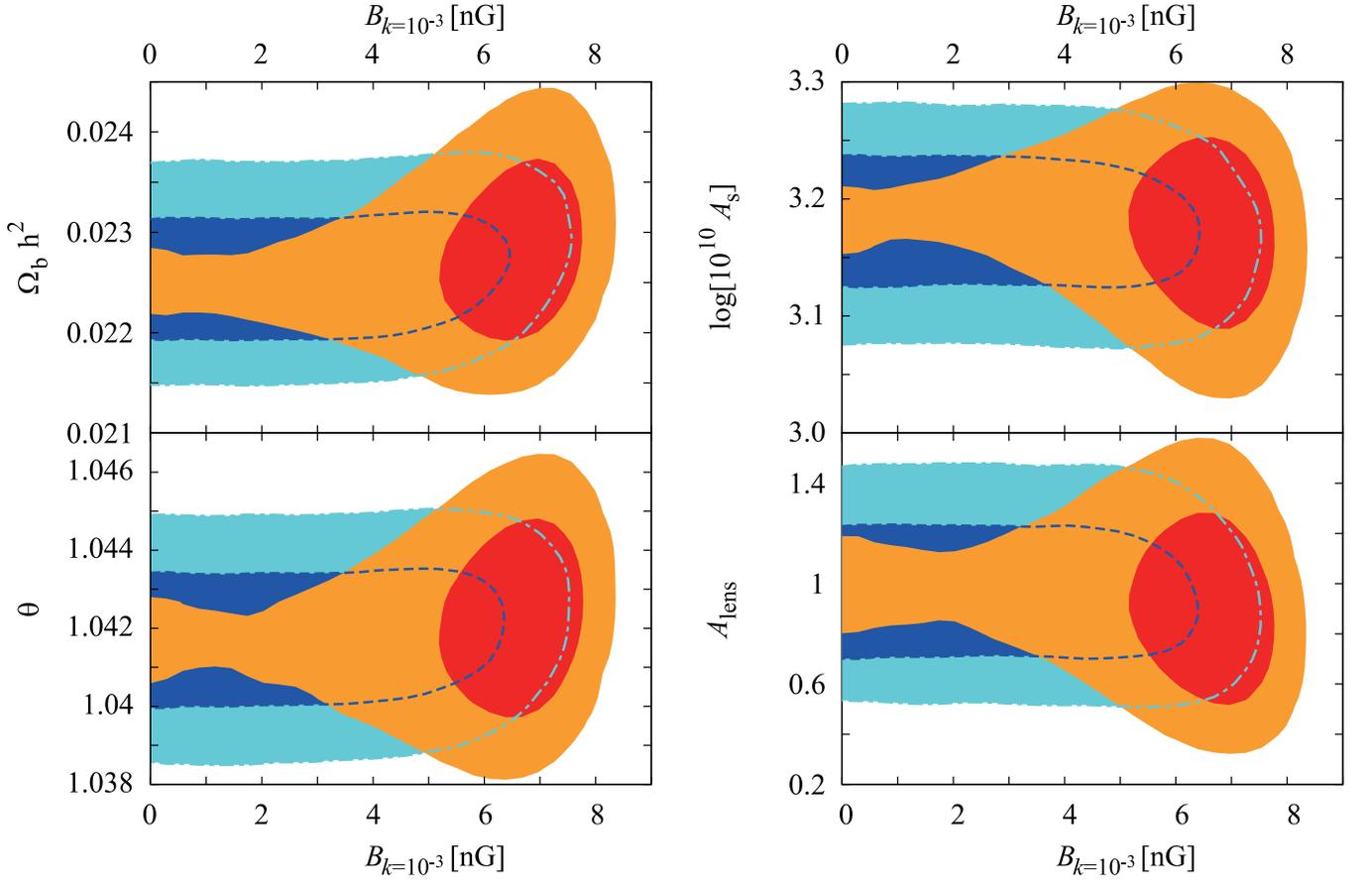}
\caption{\label{fig6}
Contours of 1$\sigma$ and 2$\sigma$ confidence limits for ($\Omega_b, \theta, A_s$, $A_\mathrm{lens}$) vs the PMF field strength $B_{k=10^{-3}}$.  
Red and orange contours show the constraint with the BB mode of the CMB, 
and blue and sky blue contours show the constraint without the BB mode.
Red and blue contours show 1 $\sigma$(68\%) confidence limits 
and orange and sky blue contours show 2 $\sigma$(95\%) confidence limits. 
} 
\end{figure}

Finally, let us argue the degeneracies between the PMF strength and other parameters. Figure \ref{fig6} shows that the PMF strength at $k = 10^{-3} \mathrm{Mpc^{-1}}$ has small degeneracies with $\Omega_\mathrm{b}$ and $\theta$. 
The amplitude around the second peak of the temperature fluctuations of the CMB decreases with increasing $\Omega_\mathrm{b}$. The amplitude on $\ell < 300$ also decreases with increasing $\theta$. On the other hand, the PMFs increase the amplitude of the temperature fluctuations of the CMB. Therefore, the effects of $\Omega_\mathrm{b}$ and $\theta$ compensate for the effects of the PMF on the temperature fluctuations of the CMB on $\ell < 300$.

We also find that the PMF strength has small negative correlations with the weak lensing effects ($A_\mathrm{len}$) and  the scalar amplitude ($A_s$) as shown in Fig.~\ref{fig6}. The amplitude of the temperature fluctuations of the CMB increases with increasing $A_s$ and the PMF strength.
In fact, from Fig. \ref{fig4}, we find that the contributions of the PMFs to the temperature fluctuations of the CMB on $\ell < 1000$ are comparable in magnitude to the contributions of the scalar amplitude ($A_s$).
The polarization isotropies of the CMB (the BB mode) also increase with increasing $A_\mathrm{len}$ and the PMF strength.
We find that the contributions of the PMF to the BB mode of the CMB on $\ell < 1000$ (Fig. \ref{fig5}) are comparable in magnitude to the contributions of the weak lensing effects ($A_\mathrm{len}$).

\section{Summary}
In this paper, we put constraints on PMFs with characteristic scales, which could be generated by causal mechanisms in the early Universe. Extending our previous work \cite{2011PhRvD..84l3006Y}, we vary the PMF strengths of five wave numbers simultaneously with standard cosmological and foreground parameters to fit the CMB and MPS data. We obtained upper bounds on the PMF strengths as 
\begin{eqnarray}
|\mathbf{B}_{k = 10^{-1}}| &<& 4.685~\mathrm{nG}~\mathrm{at}~k = 10^{-1} \mathrm{Mpc^{-1}}, \\
|\mathbf{B}_{k = 10^{-2}}| &<& 2.090~\mathrm{nG}~\mathrm{at}~k = 10^{-2} \mathrm{Mpc^{-1}}, \\
|\mathbf{B}_{k = 10^{-4}}| &<& 5.310~\mathrm{nG}~\mathrm{at}~k = 10^{-4} \mathrm{Mpc^{-1}} ~\mathrm{and} \\
|\mathbf{B}_{k = 10^{-5}}| &<& 10.91~\mathrm{nG}~\mathrm{at}~k = 10^{-5} \mathrm{Mpc^{-1}}.
\end{eqnarray}
Also we obtained a finite value for the PMF with $k=10^{-3}$Mpc$^{-1}$ as
\begin{equation}
4.867~\mathrm{nG} < |\mathbf{B}_{k = 10^{-3}}| < 7.491~\mathrm{nG}~ (1\sigma)~\mathrm{at}~k = 10^{-3} \mathrm{Mpc^{-1}}.
\end{equation}
The value of $|\mathbf{B}_{k = 10^{-3}}|$ is nonzero at more than 2.5$\sigma$ significance. This is attributed to the BB mode data from the QUAD experiment and adding the PMF at this scale reduces the total chi-squared value by about 8.3.

Since the current BB mode data have relatively large errors, we must wait for the future observations to obtain more robust conclusion. The non-Gaussianity in the fluctuations is another way to identify the PMFs. The prediction of non-Gaussianity generated by the PMF indicated in this paper will be presented elsewhere in near future.

\begin{acknowledgments}
This work has been supported in part by Grants-in-Aid for Scientific Research 
(D.G.Y. by Grant No. 25871055; K. I. by Grant No. 24340048; and K. T. by Grant No. 23740179, No. 24111710, and No. 24340048)
of the Ministry of Education, Culture, Sports,
Science and Technology of Japan.
\end{acknowledgments}

\bibliographystyle{apsrev}

\end{document}